# Certification Authority Monitored Multilevel and Stateful Policy Based Authorization in Services Oriented Grids

**Ajay Prasad** ajayprasadv@gmail.com
*School of Engineering*
*Sir Padampat Singhania University*
*Udaipur,313601, India*

**Saurabh Singh Verma** ssverma.et@mitsuniversity.ac.in
*Faculty of Engineering and Technology*
*Mody Institute of Technology and Science*
*Lakshmangarh, 332311, India*

**Ashok Kumar Sharma** aksharma.et@mitsuniversity.ac.in
*Faculty of Engineering and Technology*
*Mody Institute of Technology and Science*
*Lakshmangarh, 332311, India*

**Abstract**

Services oriented grids will be more prominent among other kinds of grids in the present distributed environments. With the advent of online government services the governmental grids will come up in huge numbers. Apart from common security issues as in other grids, the authorization in service oriented grids faces certain shortcomings and needs to be looked upon differently. The CMMS model presented here overcomes all these shortcomings and adds to the simplicity of implementation because of its tight similarities with certain government services and their functioning. The model is used to prototype a State Police Information Grid (SPIG). Small technological restructuring is required in PKIX and X.509 certificates.

## 1. INTRODUCTION

Grid Computing is a vastly distributed approach to facilitate solutions providing, resource sharing, job execution and various services. The grid term is derived from electricity grids. The analogy of electricity grid in compute grids is appropriately understood by the fact that both grids are mainly meant for supplying vastly distributed resources and services. The figures 1 and 2 depict the exact similarities between them. The Services oriented grids become more prominent as many government services are being diverted on the online platforms all over the world. The non-grid organization doesn't have connected nodes even within the organization. The user needing any service should enter the organization, go to the service site (or counter) and get the service. The organizational-bounded grids are confined to the organizational boundary. The user connects to the main node and gets its services by implicit forwarding to other internal nodes. It's understood that Virtualization is the main approach towards grids. A Virtual Organization (VO) based grids are those where services fall out of organizational boundaries, and multiple organizations jointly render services to users. The user connects to a discovery node nearby it and gets to a service node in order to get service. Figure 3 depicts organizations where some of their nodes are service nodes that take part in the grid. Grid Security has always been an issue. Providing services is not the only task that needs to be done. User's confidence and safety is a major concern and lots of energy is devoted into this factor too. Availability, integrity, confidentiality all are important and must be handled. Confidentiality consists of authentication and authorization. Authorization is to perform the necessary action to confirm that an authenticated entity is allowed to perform an action on a resource[14]. As stated in [13] there are two general approaches for authorization: identity-based or token-based. The identity based authorization mainly depends upon user identity and access control lists. On the other hand token based approaches rely on un-forgeable tokens granted by the service provider or controller which are presented by the users to avail services. Tokens of service can be passed on from one to another in form of trust management and delegation of rights. Trust[13] can be generally defined as having confidence that a party will behave in an expected manner despite the lack of ability to monitor or control that other party. In identity based authorization trust management consists of defining the sources of authorities for user identification, attribute assignment and possibly policy creation. Policies of an organization can be regarding access rights, levels of trust, whom to trust etc.

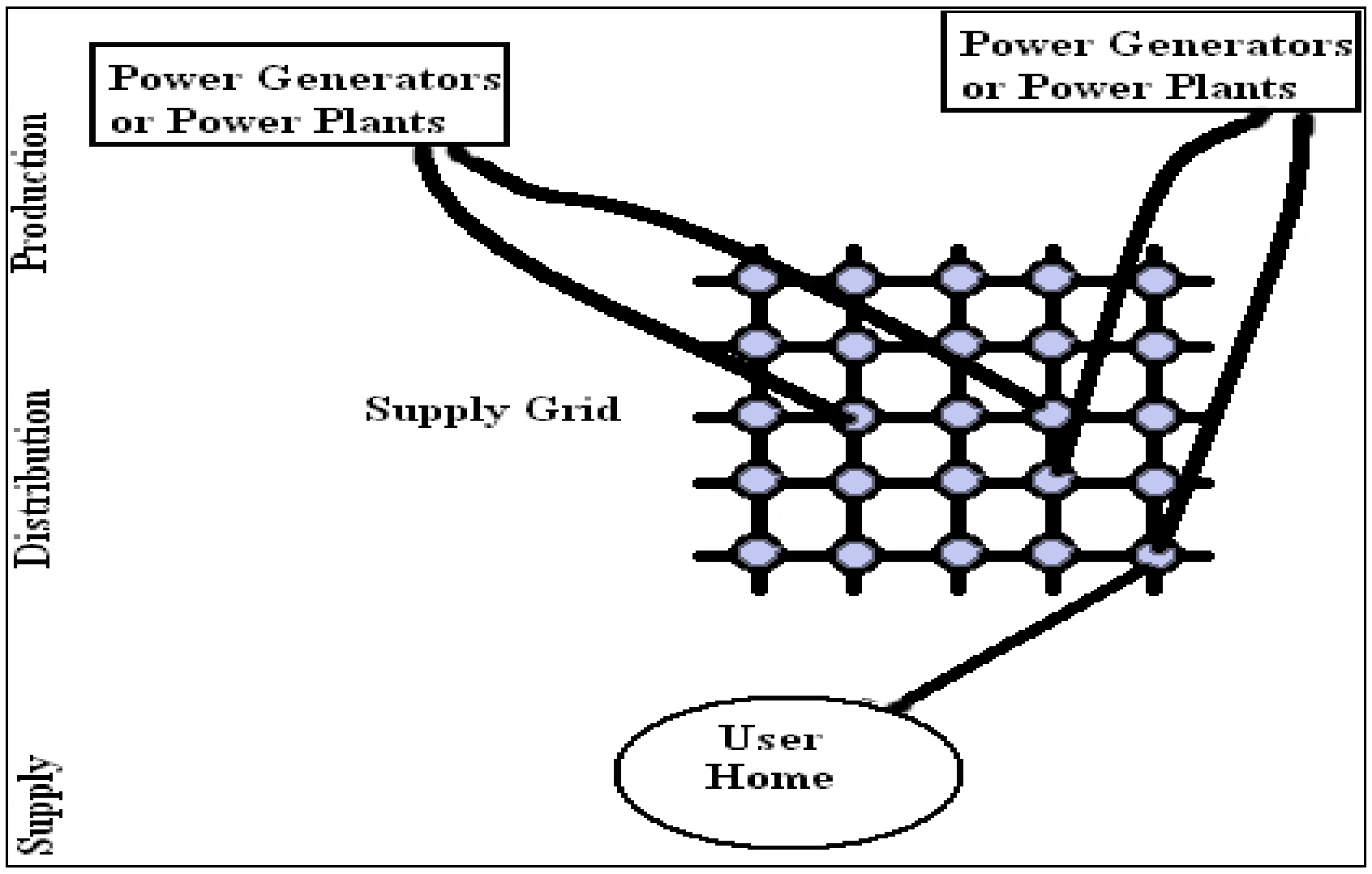

**Figure 1:** An Electricity Power Grid.

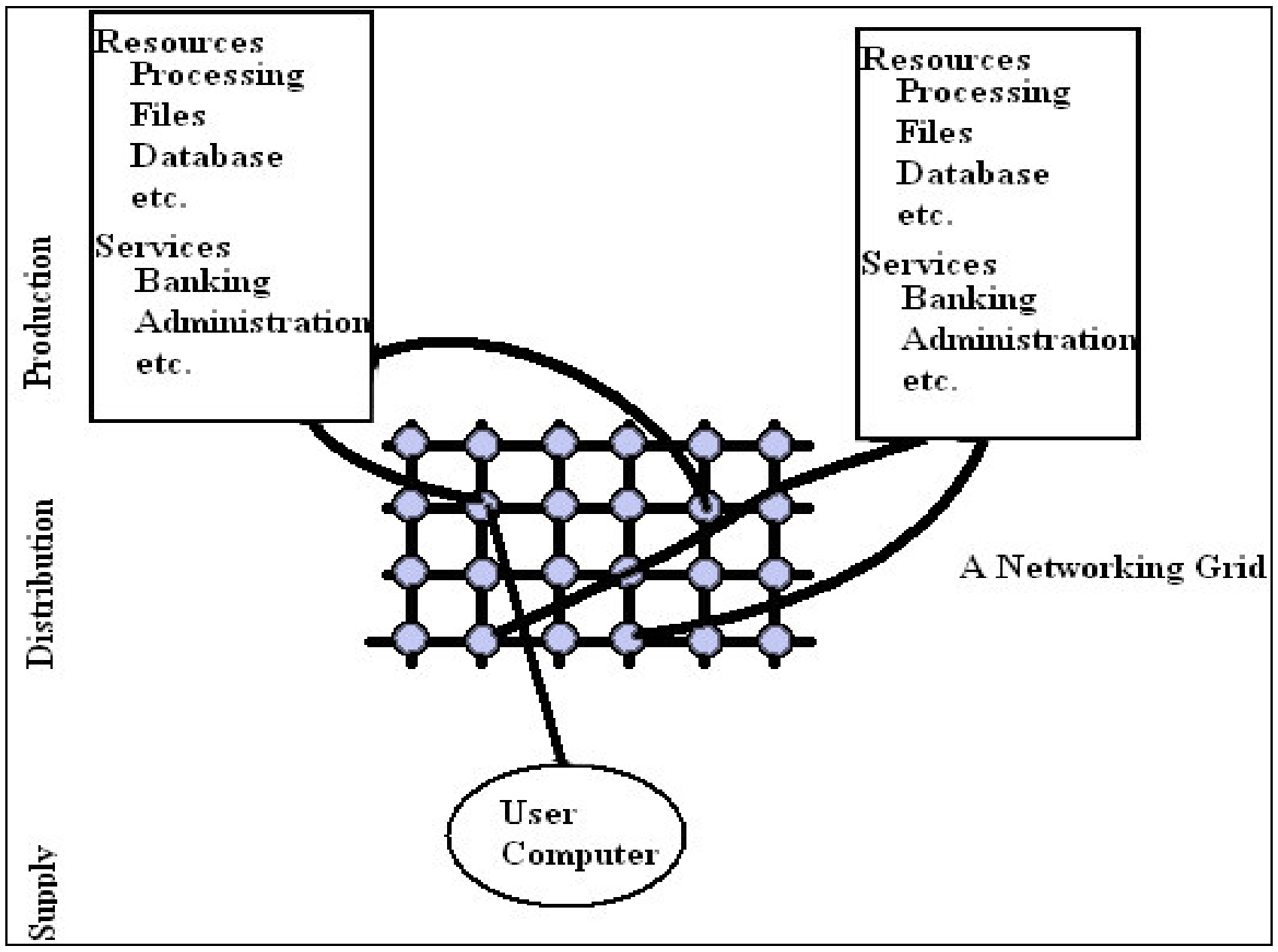

**Figure 2:** A Compute Grid.

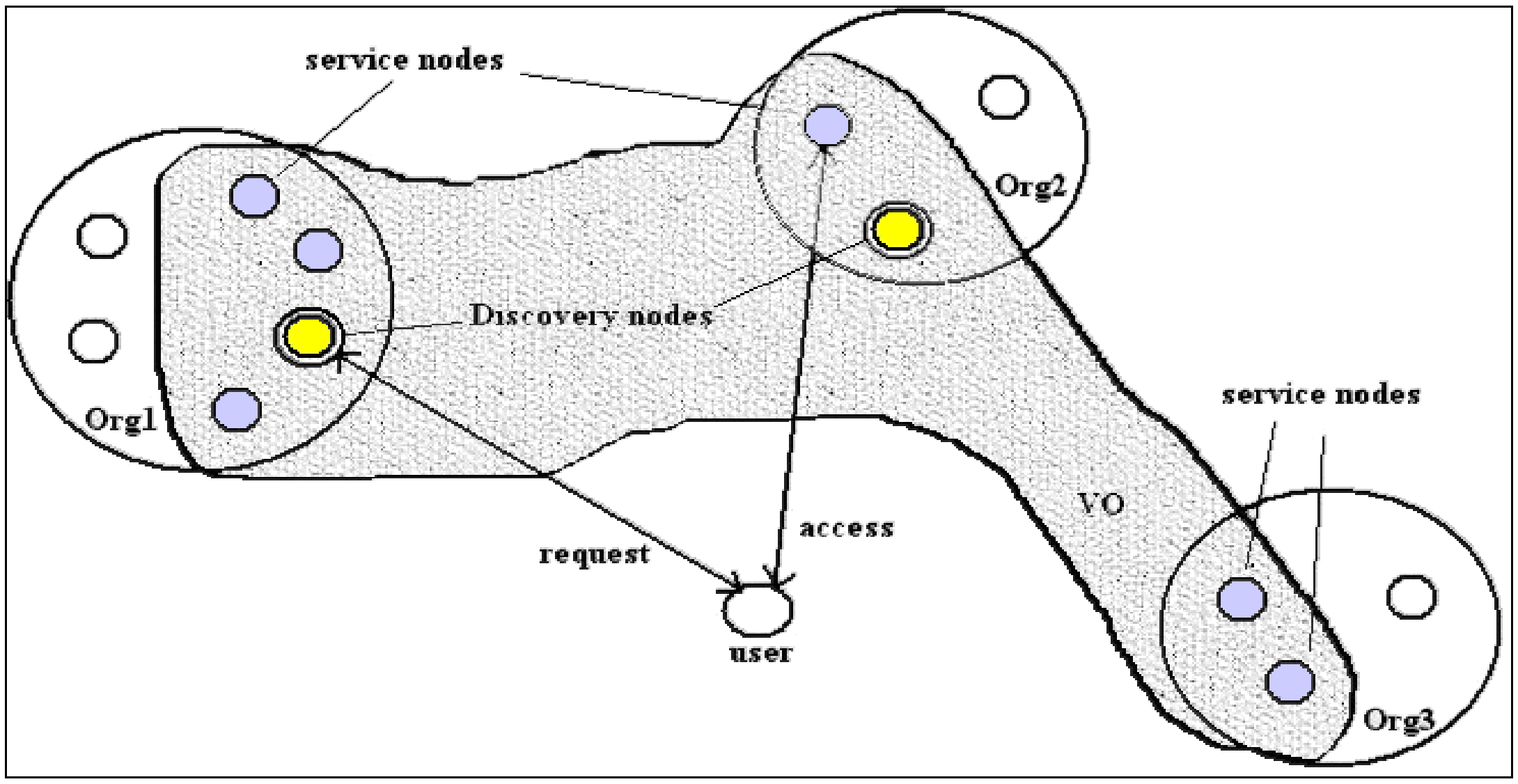

**Figure 3:** A Virtual Organization involving many organizations together.

## 2. AUTHORIZATION IN CURRENT GRID ENVIRONMENTS

Authorization has been taken care of by user identity and access control lists (distributed grid map files) in Grid Security Infrastructure (GSI)[1] incorporated in GT4.0. Also certain suggestions of a model where two points of action is required, namely, Policy Decision Point (PDP) and Policy Enforcement Point (PEP)[5]. The first one decides over the policy matters and the latter enforces the policies decided. The model can be realized both in simple and manageable centralized way or a complex but performance oriented distributed way. The governing rules regarding authorization either to grant or to deny can be acquired by the PDP in different ways[6] namely, Agent, Push and Pull models.

Trust model as suggested in [2] defines a policy database, evaluator and result of evaluation. It goes true only when dynamic, flexible and fine grained policies are maintained in a structured and simple manner. Even though dynamic policies are maintained as property based certificates in PRIMA[7], VOMS[8], CAS[9] and X.509 attribute certificates[10] there is no standard interface for using them yet. Globus Resource Acquisition and Management (GRAM) system maintains policy mappings using gatekeeper and Job manager components[12]. Akenti Policy language[11] expresses policy in XML and store in three types of signed certificates: policy certificates, use-condition certificates and Akenti attribute certificates[11].

A proposed privacy, trust and policy based authorization framework[2] introduces the concept of Filter-in and Filter-out at domains in a distributed infrastructure. With Filter-out component, the Subject leaves the Domain with access rights that his parent Domain grants to him. With Filter-in component, the Subject enters the organization with access rights that the target Domain grants to the parent Domain. In other words, the Subject gets the intersection of the rights that his parent Domain grants to him and the rights that target Domain grants to parent Domain.

A Stateful grid should maintain state information along with certification to be carried during the lifetime of the process or job or request for resource. The safety and consistency in policy based authorisation systems are discussed in [4] suggest that the collection of credentials used to satisfy a given authorization policy acts as a partial snapshot of the system within which the policy is evaluated. The correctness of an authorization decision depends on the validity and stability of the view used during policy evaluation. If we assume that each credential is stable (i.e, that the assertion stated in the credential remains true until its pre-ordained expiration time) then policy evaluation can be reduced to the problem of stable predicate evaluation on distributed snapshots[3].

## 3. SHORTCOMINGS IN AUTHORIZATION TECHNIQUES IN CURRENT GRID ENVIRONMENTS

As mentioned in the above section, authorizations in grids are incorporated or suggested in many models or papers. They all have some of the following shortcomings in them.

**Weak trust management:** In diverse and dynamic domain to domain interactions trust management helps in deciding what entities are to be trusted to do what actions [13]. The trust management itself is a weakness in grid environments. However, large distributions of services have to rely on trust to some extent. Trust based on mere confidence can lead resources and services in wrong and malicious hands and can corrupt vital resources and their accessing.

**Policies are static:** Domain specific policies are essential so as to integrate multiple policies and varied policies of entities providing services and resources in a grid. Currently users must keep track of the right/required credential for each resource they might access.

**Multi-nodal policies are not considered:** Each entity or domain in a grid can be seen as a node, every node can have its own policies. Since a request can be serviced by combination of several nodes, their composite policies have to be considered or at least the requester must fulfil policy requirement of all the nodes once before getting started. This filtering is required from both sides (the requester and the service provider). This scenario can occur and can be fatal in case of job submissions in a grid (processing grids). If a job is submitted to a grid and one or more nodes later on reject one or more sub-jobs at the time of join then it could cause unnecessary delay and even losses.

**State information of processes and/or users are not included in the authentication procedure or authentication exchanges:** Keeping state information in authentication dialogues will help the policy management (supporting dynamic policies) to perform policy mappings[14] in more secure manner. In addition to the processes/requests the above goes true for any role based application working on a VO. The Dynamic Role Based Access control in SESAME[16] suggests a model but it is confined to role based access only.

**Coarse grained policies and services:** Coarse grained policies[14] results from the common reliance on standard operating system enforcement mechanisms. These are constrained by expressiveness limitations. It also poses an accounting problem if the user employs resources in conjunction with separate projects or applications in separate VOs. Coarse grained services result in scalability problems and need complex mechanisms for policy mappings.

## 4. THE CMMS MODEL

At the outset we need to take into consideration the scenario where the proposed model will work and then things can be placed in their respective slots. The service oriented grid is a grid where user can plug in (or connect) to get desired service (only if they are authorized to). One node can also act as a user while forwarding a request to the next node. The Grid Security Infrastructure's delegation of rights and negotiation of trust[15,17] can be performed in these cases. When a user connects to the VO discovery node, the discovery node performs initial checks and looks for a free or available service node and sends the address of the node to the user. Let us list out all the players in this grid scenario as shown in figure 4.

1. Users.
2. VO discovery nodes.
3. VO service nodes.
4. Certification authority.

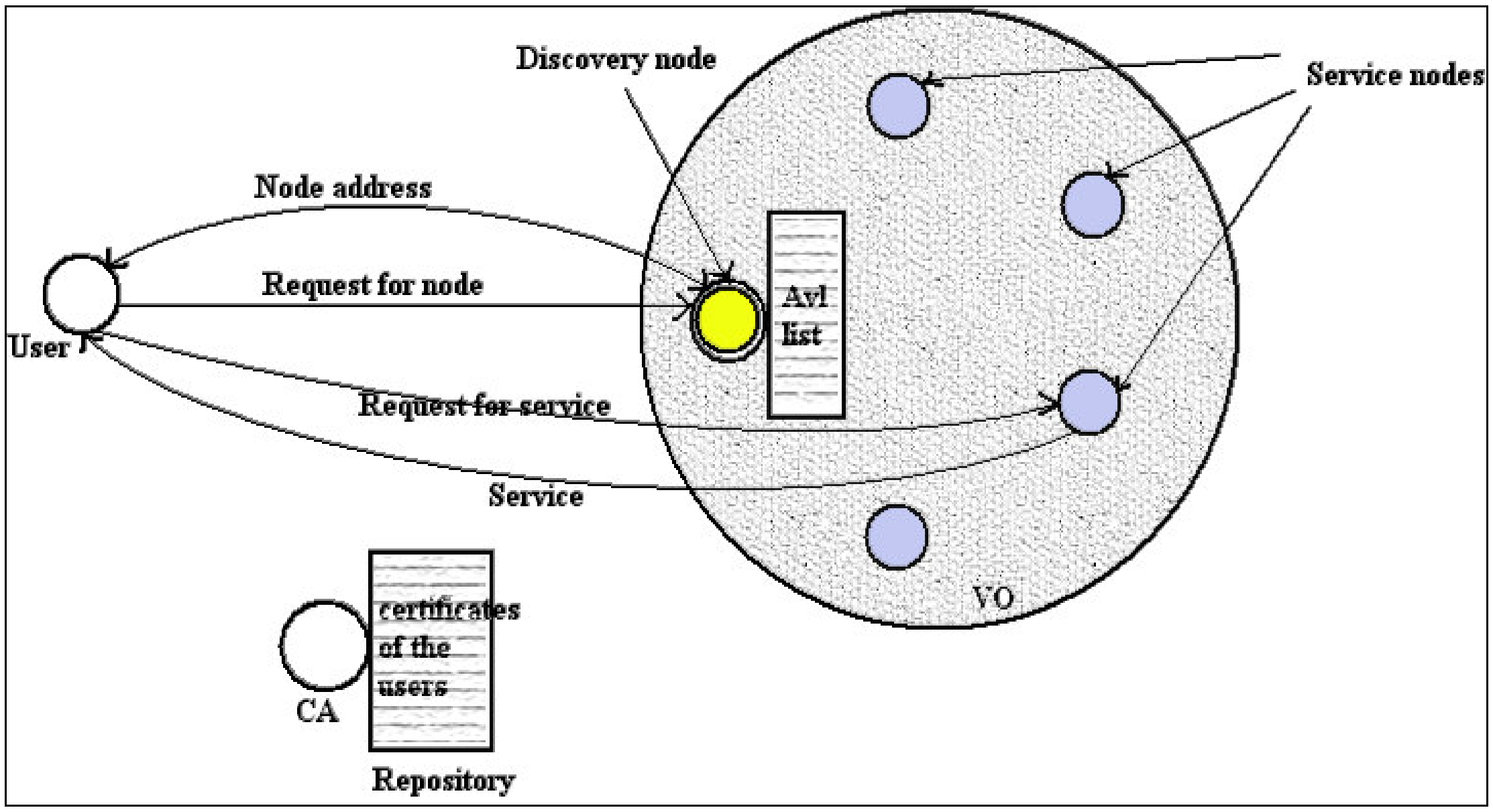

**Figure 4:** A normal service oriented grid scenario.

The users are the ones who may need to get the services of any node in the VO. The VO discovery nodes are responsible to perform initial authentication and give the access to an available node (address). The users then connect to the service node directly and get the required service. In all these dialogues the important aspect of authentication is taken care of by the certificates maintained by CA.

The proposed model will be consisting of four set of applications apart from general middleware at grid, CA and user site. Let's discuss these applications as we come across them one after another. The modified parts are put into the grid scenario in the figure 4 as figure 5. Major Players in the scenario are:

1. User application
2. Discovery node
3. Service node
4. CA node
5. Repository node
6. Monitor node

The user has an application in its system. The application will be basically of two parts or modules:

1. Get access part
2. Services interface part.

The get access part is responsible to perform connections and authentications and request for authentications both to VO discovery node and to the services nodes. This part is also responsible for requesting certificates of VO discovery nodes and service nodes to start authentication process. The get access part on receiving the service list activates the service interface part. Only those interfaces are initialised and activated which are in the service list that is sent by the service node. In case of forwarding of request (delegation of rights) the service list have to be matched at the next node and policy will be mapped once again and services need to be rendered if the mapping too has the user authorized for the service that has desired the request forwarding.

The discovery node is responsible to perform level1 authentication. It will perform Filter and Impose on the user states in the certificates. The discovery node maintains a list of available or free nodes. The discovery node selects one free node and sends it back to the user node.

The node address is sent to the user and the user get-access module connects to the service node. The service node contains two modules:

1. Authentication and Authorization module.
2. Service renderer module.

The Authentication and Authorization module is responsible to perform level 2 authentication demanding user's certificate from CA repository node and then perform policy mapping from the user's effective state sent by the discovery node. The policy mapping is explained in the later section.

The monitor node performs state or policy changes for users and service nodes. It sends signals to CA for policy changes and CA writes new certificate and sends it to the repository node which performs store cert to store it. The service renderer module will also perform service request forwarding and delegation of rights for the user.

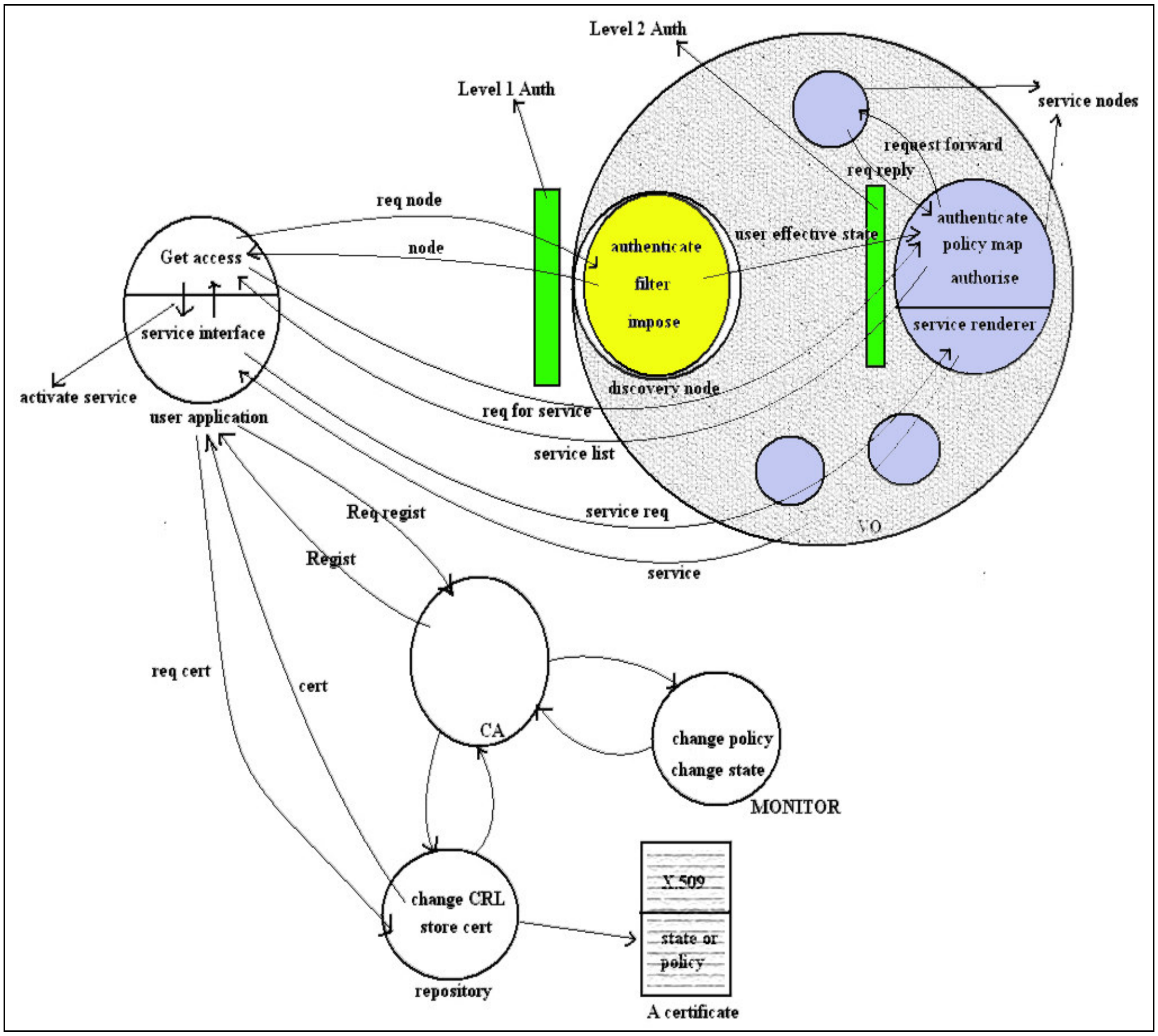

**Figure 5:** The CMMS overall scenario

Let us now have a look at the complete flow of the scenario as depicted in figure 5.

1. REG_USER: The user registers with the CA. The CA consults the monitor for initialising the state of the user.

2. REG_DISC, REG_SERV: The VO discovery node and service nodes are all certified similarly.

3. The user now installs the user module and invokes it and connects to the discovery node.

4. GET_CERT: The user requests CA repository for discovery node certificate, verifies it and extracts the node's public key and sends GET_NODE to discovery node.

5. The discovery node requests CA repository for the user's certificate, verifies it and extracts user's public key. It then performs AUTHENTICATE, FILTER and IMPOSE.

6. SEND_NODE: The discovery node selects a free service node and sends its address to the user.

7. SEND_EFF_STATE: The discovery node simultaneously sends the effective state of the user with the user name to the service node.

8. SERV_REQ: The user on receiving the service node address connects to the service node and sends service request message to the service node.

9. The service node on receiving the SERV_REQ from the user, requests for its certificate (GET_CERT) and verifies it, extracts user's public key.

10. Then the service node performs POLICY_MAP and sends the authorised SERVICE_LIST to the user.
11. The user avails the service.

(All the communications are through X.509 authentication procedures. The SPIG implementation has not implemented the authentication procedure as its just a prototype.)

***The Middleware architecture:***

The overall model can be viewed in an architectural fashion suited for implementations. The figure 6 depicts the modules that must be included as per incorporating the CMMS model. The figure is quite self explanatory. and has not been explained in this paper.

***X.509 certificates and PKIX***

The PKIX components are added with a Monitor node and every X.509 certificate will hold extra space for keeping the policies and states of users and service node. The space either can be adjusted from the v3 extensions in X.509 or can be added as required. The adding of an extra component in PKIX and an extra space in the X.509 format will never affect the authentication procedure of X.509 at all.

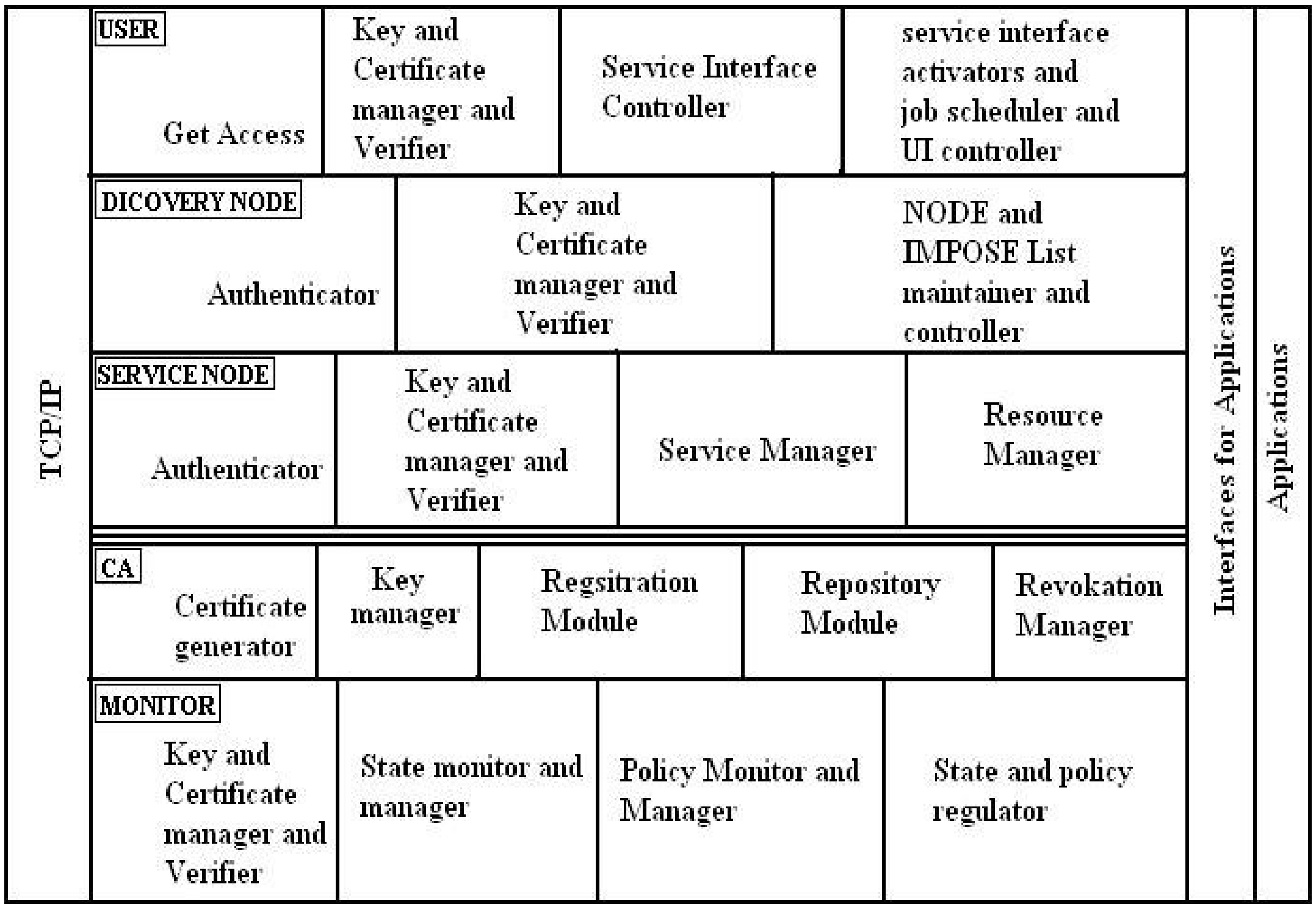

**Figure 6:** The CMMS Middleware Architecture.

## 5. EXAMPLE: STATE POLICE INFORMATION GRID (SPIG) PROTOTYPE IMPLEMENTATION

The concept of CMMS is presented hereby through a prototype implementation called as State Police Information Grid (SPIG). The grid prototype is a service oriented VO keeping in mind the present Indian Police Department Infrastructure. The police in India are structured as police stations covering a specific area of control and authority. Every police station is responsible to take care or ensure law and order in their specified region. The activities of every police station are taken care of by number of officials ranging from Station in-charge to a set of inspectors. The groups of police stations are under the domain of Superintendents, Inspector generals etc. The overall policing of a state are governed by the state Home Ministry. The officials of Home Ministry range from political leaders to a set of IAS officers. The Home Ministry is responsible to monitor the police activities, regulate rules and policies, assign roles and perform authorized activities like transfers, suspension etc. The whole structure of policing as explained above might not be exact to the existing structure but that's not important. What's important for us is to understand the essence of CMMS with the structure.

In a nutshell the whole structure consists of following components.

1. State Police.
2. Police stations.
3. Home Ministry.
4. Officials of Home Ministry.
5. Officials of police.
6. Citizen of the state.

The whole CMMS can be realized with these components very simply. Lets map all the above components to the components of CMMS.

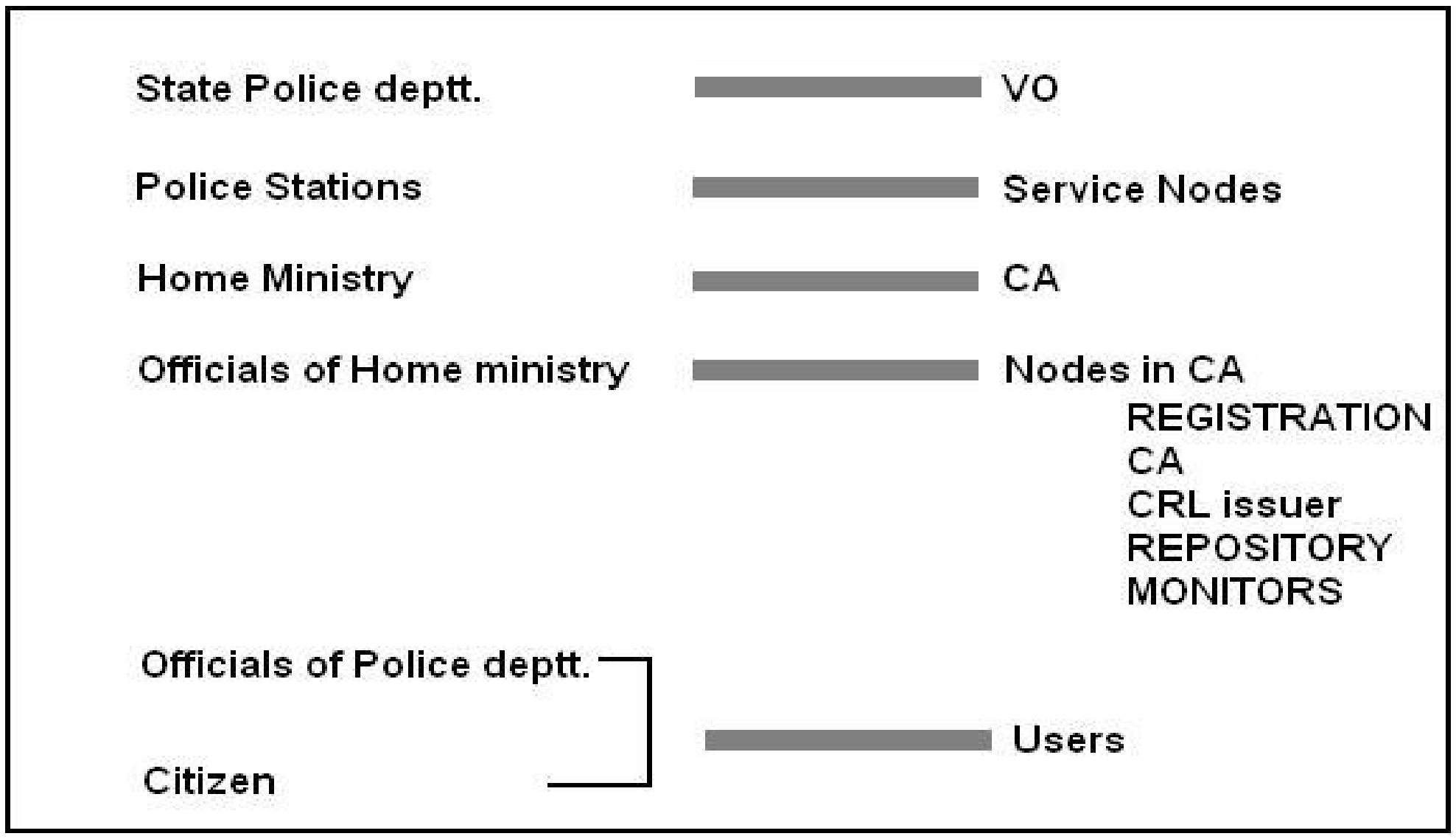

**Figure 7:** State police structure mapped to the CMMS

The whole prototype is actually a set of clients and servers. The list below is the clients and servers implemented. They are not exactly the same number of servers as we'll have in CMMS. The later figure shows the mapping which reveals that actually the roles and tasks required in CMMS are all covered in the implemented set of servers.

The List of client and servers are:

1. USER
2. VO_SERV_NODE (SPIG)
3. VO_DISC_NODE (SPIG)
4. CA _NODE (CA)
5. CA _CRL_REP_NODE (CA)
6. CA _MONITOR_NODE (CA)

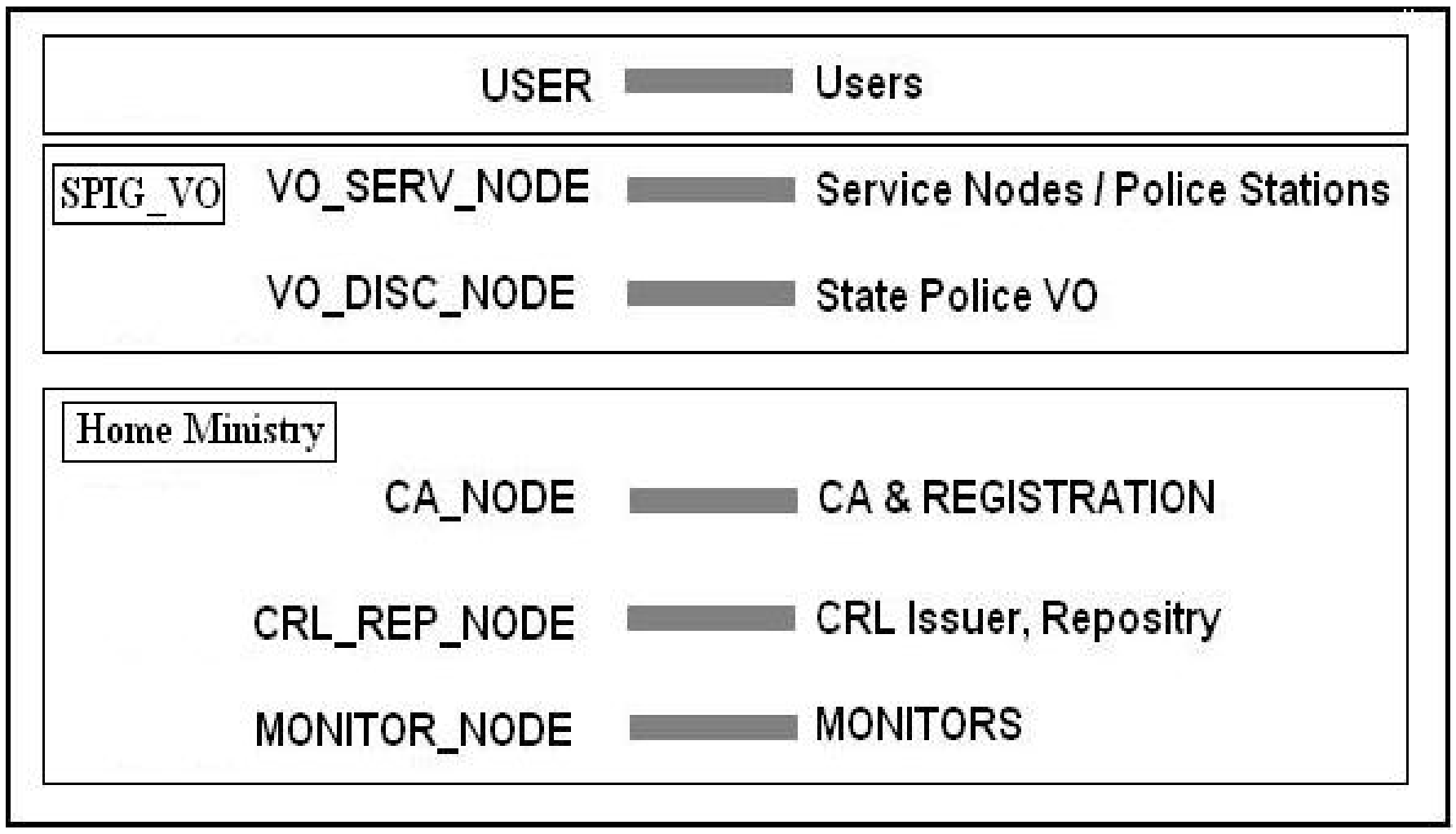

**Figure 8:** The implemented servers and their CMMS counterparts.

## 6. STATE TO POLICIES AND POLICIES TO SERVICES MAPPING IN CMMS

### 1) State in CMMS:

In most of the earlier authorization in grids more emphasis was given to roles and privileges. But, a stateful grid must take states of its user as well as servicing nodes into consideration while authorization. Here are few reasons why states are important in grid authorization.

1. Both users and the service nodes are subject to change (for users it can be personality or position and for service nodes it can be load and vulnerability).
2. Different users might have different policies as to how and when they'd like to take help from.
3. Different servicing nodes might have different policies regarding rendering of services to a user in a particular state.
4. It is important that the audit records of the services or resources are maintained as to which user and in what various states they have been used.
5. In case of processing-grids process state play roles while joining or forking a process over multiple nodes.
6. Suspicious nodes or users can be traced out and stabilized.

***Example: states in the SPIG prototype.***

| State | Ref No |
|---|---|
| On Duty | 1 |
| Suspended | 2 |
| Transferred | 3 |
| Convicted | 4 |
| On Leave | 5 |
| View Restricted | 6 |
| Edit Restricted | 7 |
| User | 8 |

**Table 1:** States of users used in SPIG.

A user state can change and in that case many accesses he/she was authorized may become vulnerable and might require restriction imposed on the accessibility of the user. The states of a user are a fact due to some environmental or situational changes but are very important in authorization. As shown in the table 1, The states 1 to 7 are for State Police employees or officials. The states 4, 7 are for common man or citizen of state.

In the same manner the states of the VO and its service nodes are also very important (The VO and Service node states are not included in SPIG prototype). For example if a VO is a disowned company then the users might not like to use it in full confidence. If a service node is in a state of partial failure or under control of certain enemy organization, it will not be encouraging for the users to use it for any purpose. However, considering the major issues the user state is vital and has been touched in SPIG implementation.

The state structure can be as:

**<state-list> -> {<state>, <state>, …}**
**<state> -> 1/2/…/n** *for n numbers of states maintained.*

One can be having more than one state.

**2) Filter and Impose at VO level:**

**Filtering:** Certain states of user which are not needed in policy mappings in a VO needs to be separated from the set of states in the user certificates.

**Imposing:** The discovery node will maintain a IMPOSED LIST and will add/impose certain other states to the user based on the list on per user basis.

***Example: Filtered States***

| States | States considered at present | Filtered states |
|---|---|---|
| 1, 4,15 | 1,2,3,4,5,6,7,8,9,10,11,12 | 1, 4 |

**Table 2:** Example: Filtered states.

State 15 is not under consideration in the VO so only state 1, 4 will be used in mapping.

***Example: imposed States***

| In user certificate | Filtered state | Imposed state | Effective state |
|---|---|---|---|
| 1, 4, 15 | 1, 4 | 11, 12 | 1, 4, 11, 12 |

**Table 3:** Example: Imposed States.

VO can impose certain states to certain users as per the IMPOSED_LIST maintained at VO level. The states imposed will vary from user to user and can be dynamically maintained at VO level in a IMPOSED_LIST. The imposed states can be determined by certain management policies which may or may not be programmed.

**3) Services:**

Services can be numbered in a very fine grained way so that they can be rendered that way. That is, a service like Police verification can be read as well as upgraded and applied for. These can be given unique numbers as:

1. Police verification: search and read
2. Police verification: search an upgrade
3. Police verification: apply

With fine grained services and policies and states are can have better control over authentication and less vulnerable to attacks.

***Example: Services in SPIG.***

| Services | ref no. |
|---|---|
| Criminal Records Database | 1 |
| FIR Records | 2 |
| Search for INV status | 3 |
| ADD FIR/Criminal Records | 4 |

**Table 4:** Services Provided in SPIG.

**4) Policies.**

Policies are like rules to be followed in order to allow or disallow a user from getting a set of services and also accepting or rejecting a service from a node by the user. Generally policies are based on states like for example a typical policy can be:

7: 3,4

That is, the users having state 7 will be given service number 3 and 4 only.
There can be several policies for various permutations of states. The policy structure can be as:

**<policy> -> {<state> : <service-list>}**
**<service-list> -> {<service>, <service>, …}**
**<service> -> 1/2/3…/m**
m= number of service rendered at that node

**Example: A policy used in SPIG.**

{1: 1, 2, 3, 4}
{2: 2, 3}
{3: 2, 3, 4}
{4: }
{5: 2, 3, 4}
{6: 2}
{7: 2, 3}
{8: 2}

**Table 5:** Example: Policy of a node

**5) State-Policy to service mapping.**

Based on the policies and user's state the service nodes can decide as to which services it will provide to the user. At the VO discovery node effective user state will be packaged and sent to the service node for mapping.

A typical procedure of mapping at service nodes is explained below. The procedure is used in the SPIG implementation. The service node will tally the effective state sent by the VO level to the state in the user certificate and perform mapping.

Lets see how the policies can be represented. The SPIG uses an exact representation as shown below:

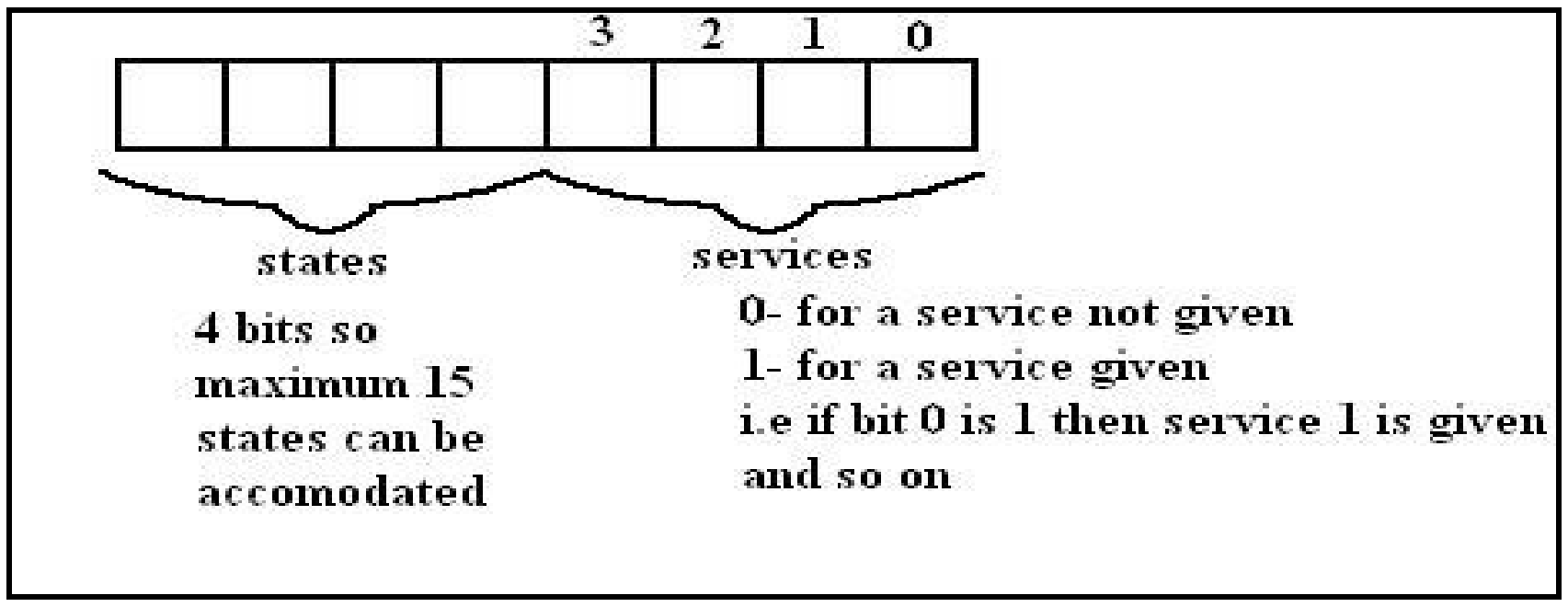

**Figure 9a**: Representing Policies

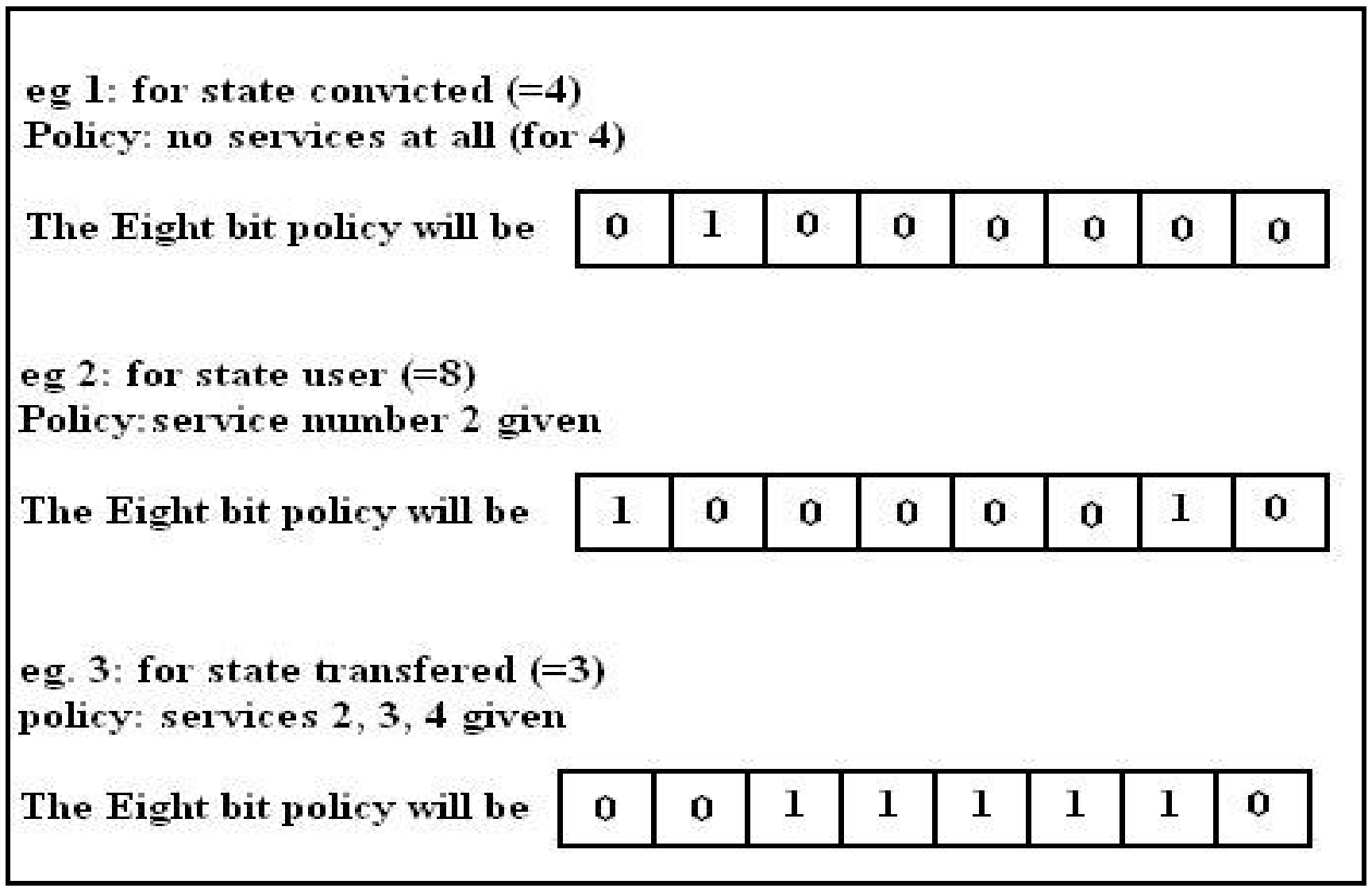

**Figure 9b**: Example: keeping state and policy

Now, the mapping will be as below (using the policy stated earlier)

Say if a user has the state 5, 6
The mapping will be performed as follows

Goto $5^{th}$ policy (94) = p1 = **01011110**
Goto $6^{th}$ policy (98) = p2 = **01100010**

By masking get service for $5^{th}$ policy

```
          01011110
    AND   00001111
          -------------
    Service1 00001110 =14
                  means given services 2,3 and 4
```

Similarily

```
    Serivce2 00000010 =2
                  means given service 2
```

Now, since there are more than one set of mapped services only those services will be rendered which are common in both the sets. Therefore we will AND both these services.

```
ANDing    00001110
    AND   00000010
          ------------
          00000010 = 2
```

Therefore the mapped service is = 2
The user will thus get only $2^{nd}$ service.

The same mapping technique is exclusively followed in SPIG implementation. Here is the working scenario in CMMS as implemented in SPIG. The numbers in the figure 10 depict the sequence in which the request or response will take place.

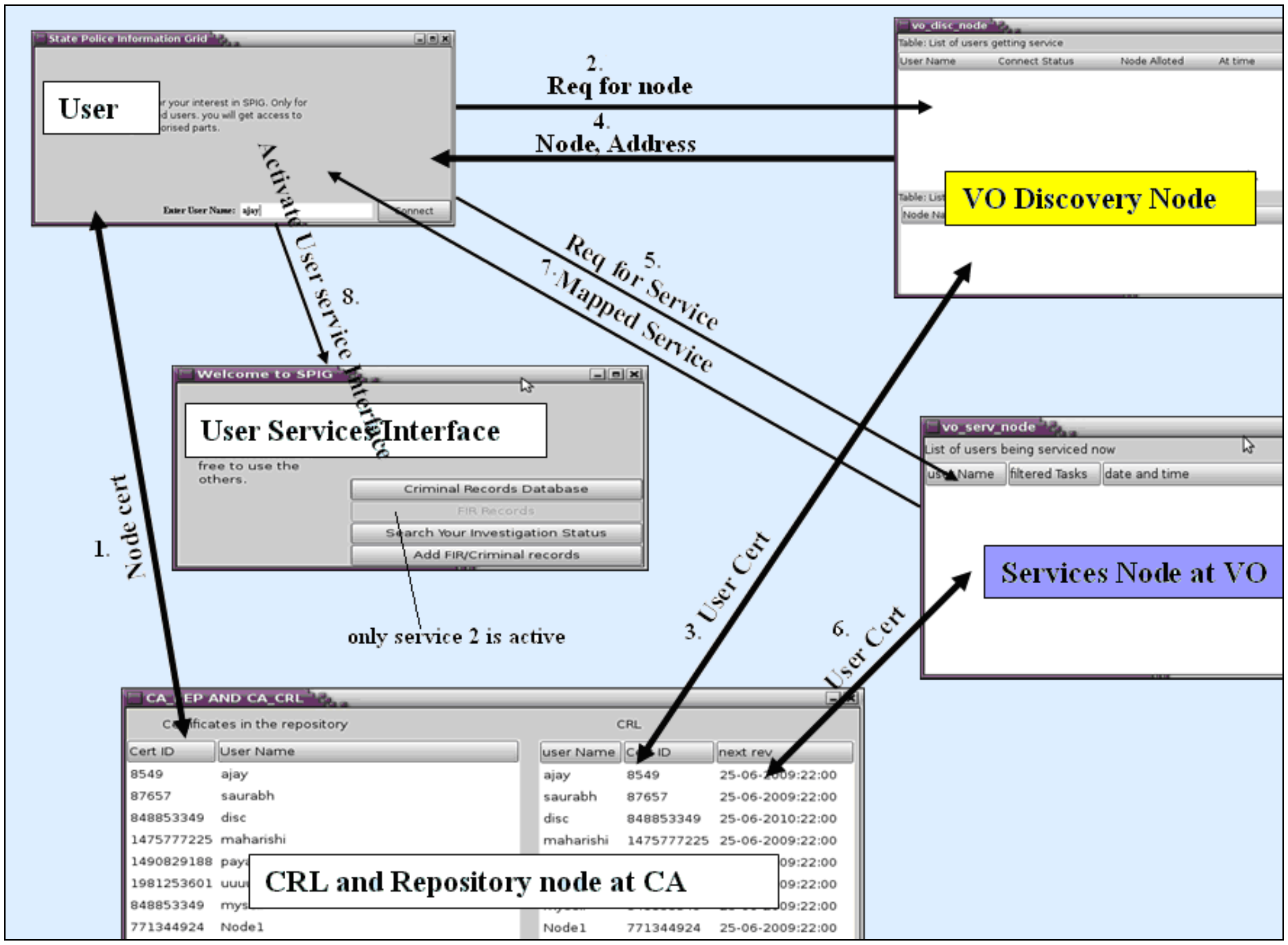

**Figure 10:** A SPIG(CMMS) working scenario.

## 7. Conclusion

Grid is a useful and exciting way of establishing large scale distributed and sharing of data, resources and services. Service oriented grid and VO for service rendering are also of prime importance as any other grid solution. Providing large scale distributed services requires high level of authentication and authorization. Existing grid infrastructures rely mainly on PKI and GSI. The available authorization by use of PEP or PDP, Push-Pull models are not addressing all shortcomings in grid authorization namely lack of statefulness, fine grained policies and services, multi nodal policies etc. Though many researchers pointed out suggestions, a combined solution for any infrastructure is yet to come out. The CMMS suggests multilevel authorization, one at VO level and one at service level. Apart from this it introduces the concept of states to both user's, VO and service nodes. Although it suggests one way state-policy to service mapping. The two-way model is very similar and can be framed very easily. The proposed model requires certain technological changes in PKI and X.509. Mainly states needs to be monitored through MONITOR nodes in CA and X.509 certificates should contain state or policy or both together. SPIG is a prototype implementation to understand the essence of the whole proposed CMMS model in a very practical grid services scenario.

## 8. Acknowledgment

Our sincere thanks to all friends and colleagues who supported us at each respect and showed us our strengths and weaknesses. Lastly, heart-full thanks to Payal and Rimi.